\documentclass[preprint,10pt,a4paper]{aastex}
\usepackage[colorlinks,urlcolor=blue,citecolor=blue,linkcolor=blue]{hyperref}
\usepackage{amsmath}
\usepackage{graphicx}
\usepackage{multicol}
\newcommand{\be}{\begin{equation}} 
\newcommand{\ee}{\end{equation}}
\newcommand{\nn}{\mbox{} \nonumber \\ \mbox{} }
\newcommand{\ba}{\begin{eqnarray}}
\newcommand{\ea}{\end{eqnarray}}

\newcommand\eg{\textit{e.g.,\ }}

\newcommand{\NS}{neutron star}
\newcommand{\NSs}{{neutron stars}}

\newcommand{\Lf}{Lorentz factor}

\begin{document}

\title{Conditions for jet break-out  in neutron stars' mergers}

\author{Maxim Lyutikov\\
Department of Physics and Astronomy, Purdue University, 
 525 Northwestern Avenue,
West Lafayette, IN
47907-2036 }

\begin{abstract}
We consider conditions for jet break-out through ejecta following mergers of  \NSs\ and provide simple  relations for the break out conditions. We demonstrate that: (i) break-out requires that the isotropic-equivalent jet energy $E_j$  exceeds the ejecta  energy  $E_{ej}$  by
$E_j \geq E_{ej}/ \beta_0$, where $ \beta_0 = V_{ej}/c$,  $V_{ej}$ is the maximum velocity of the ejecta. If the central engine terminates  before the break out, 
the shock approaches the edge of the ejecta slowly $\propto 1/t$; late break out occurs   only if at the termination moment the head of the jet  was relatively close to the edge.
(ii)  If there is a substantial delay between  the ejecta's and the jet's launching,  the requirement on the jet power increases. (iii)  The forward shock driven by the jet is mildly strong, with Mach number $M\approx 5/4$ (increasing with time delay $t_d$); (iii) the delay  time  $t_d$ between the ejecta and the jet's   launching  is important for $t_d > t_0= ({3 }/{16} ) {c M_{ej} V_{ej}}/{L_j} = 1.01 {\rm sec} M_{ej, -2} L_{j, 51} ^{-1} \left ( {\beta_{ej}} /{0.3} \right)$, where $M_{ej}$ is ejecta mass, $L_j$ is the  jet luminosity (isotropic equivalent). For small delays,  $t_0 $   is also an estimate of the break-out time. 
 \end{abstract}

\section{Introduction}

During  mergers of neutron stars \citep[NSs,][]{2017ApJ...848L..13A,2017ApJ...848L..14G}   a dense wind   is ejected \citep[\eg][]{2017ApJ...848L...6L,2018ApJ...852L..30P,2018MNRAS.479..588G}. Mass of the wind is $\sim$ few $  10^{-2} M_\odot$ and the typical velocity is a fraction of  the speed of light \citep{2010MNRAS.406.2650M,2011NewA...16...46B,2017Sci...358.1559K}.

At the same time  the tidally stripped material forms an accretion disk that feeds the newly formed black hole (BH) with magnetic flux. After sufficient amount of the magnetic flux is accumulated, the Blandford-Znajek mechanism \citep{BlandfordZnajek} leads to jet  launching, possibly with considerable delay \citep{2011MNRAS.417.2161B,2018ApJ...852L..30P}.
As a result, the jet has to plow through the expanding ejecta. Depending on the parameters  the jet  may break out, or ``fail'' -  just dissipate its energy within the ejecta
\citep{2018ApJ...866....3D}. 

In this paper we consider the jet dynamics within an expanding ejecta and formulate criteria for the jet break out. 
There is  number of numerical simulations of the problem \citep[\eg][]{2005A&A...436..273A,2018ApJ...866....3D,2018MNRAS.479..588G,2019arXiv190905867H}. Yet  high computational costs involved in the simulations often preclude a detailed investigation of the parameter space, \eg\ the dependance of the jet dynamics on the properties of the ejecta, like maximal velocity (see, though a comment after Eq. (\ref{td})).

\section{Constant driving}

We assume that the ejecta expands homologusly, with $v \propto r$, with constant proper density
\ba &&
\rho _{ej}= \frac{3}{4 \pi} \frac{M_{ej} } { (V_{ej} t)^3} 
\nn &&
E_{ej} = \frac{3}{10} M_{ej} V_{ej}^2,
\nn &&
v_r = \frac{r}{t},\, r \leq V_{ej} t
\ea
where $M_{ej}$ and $\rho_{ej} $ are total mass and density of the ejecta and $V_{ej}$ is the maximal velocity;
a more general scaling of $\rho$ can also be used, $\rho \propto  t^{-3} f(r/t)$, $v_r \propto(r/t) f(r/t)$, see   \cite{Chevalier82} and Appendix \ref{profile}. We assume that the maximal ejecta velocity is mildly relativistic at most, $\beta_{ej} =  V_{ej}/c \leq 1$.

Consider first the case when  a light, relativistic, constant power jet is launched into expanding ejecta by the central BH.
To model the advancement of the jet's head we use  the  Kompaneets approximation \citep{1960SPhD....5...46K,1995RvMP...67..661B}.  The  Kompaneets approximation
assume pressure balance between the jet and the ram pressure of the ejecta.  It requires that the contact discontinuity is close to both the forwards shock in the ejecta and the termination shock in the jet. Under certain conditions it provides an excellent  simple approximation \citep[\eg][]{1995ApJ...448L.105M,2002MNRAS.337.1349R,2003MNRAS.345..575M,2011ApJ...740..100B}. 

We assume that the jet is launched with highly relativistic velocity, $v_j \approx c$. This requires jet to be sufficiently light. The composition of the jet is not important - only its' total power (or, in fact, thrust). On the other hand, we assume that the advancement of the jet's head is non-relativistic. This is justified since both the ejecta's velocity and the jets' head are expected to be mildly relativistic  at most \citep[for relativistic treatment see ][]{2012MNRAS.421..522L}.

The Kompaneets approximation aims to capture the overall dynamics of the shock, neglecting the details of the subsonic parts of the flow, \eg\ the formation of the cocoon \citep[\eg][]{1998MNRAS.297.1087K,2007ApJ...665..569M}.   In fact, the Kompaneets approximation is momentum, not energy conserving.  In what follows, 
 we assume that the dynamics of the head of the jet  is also self-similar, \eg\ no sideways expansion. This is justified   if  the properties of the jet at launching remain constant (\eg same opening angle).    

In the  Kompaneets approximation the shock radius  $R(t) $  evolves according to (see Appendix \ref{relatKomp})
\be 
\frac{L_j}{4 \pi R^2 c} = \rho_{ej} ( \dot{R} - R/t)^2
\label{Komp1}
\ee
Eq. (\ref{Komp1}) expresses a balance between the pressure of the relativistic jet and the ram pressure of the expanding ejecta. 
It is a generalization of the often-used momentum balance equation \citep[eg][their Eq. (1)]{1989ApJ...345L..21B},  to expanding and time-dependent external medium 
\cite[see also][]{2011ApJ...740..100B,2003MNRAS.345..575M}. Relation (\ref{Komp1}) supersedes (in the relevant  non-relativistic regimes)  the related instantaneous approximation for velocity used in other works
\citep[\eg][]{2018ApJ...866L..16M,2019arXiv190707599S,2019arXiv190905867H,2019ApJ...876..139G}.

If the jet is launched with a  delay $t_d$, the solution is
\be
R=  t \left(  t^{1/2} -t_d^{1/2} \right)^{1/2}  \frac{2 L_j^{1/4} V_{ej}^{3/4}}{ 3^{1/4} c^{1/4} M_{ej}^{1/4}} , \, t \geq t_d
\label{Roft}
\ee
 $L_j$ is the isotropic equivalent jet power,
  $t_d$ is delay time
  between the onset of the ejecta and  switching on jet. 
  
   In Eq. (\ref{Roft}) 
 time $t=0$ corresponds to the initial explosion. Shifting time to the moment the jet is initiated, we find the evolution equation for   the expansion of the jet-driven bubble
\be
R= ( t+t_d)  \left(  (t + t_d) ^{1/2} -t_d^{1/2} \right)^{1/2}  \frac{2 L_j^{1/4} V_{ej}^{3/4}}{ 3^{1/4} c^{1/4} M_{ej}^{1/4}}  , \, t \geq 0
\label{RR}
\ee
For  very small delays $t _d \rightarrow 0$  Eq. (\ref{RR}) simplifies  
\be
R=  \frac{2}{ 3^{1/4}}  \frac{ L_j^{1/4} V_{ej}^{3/4}}{ c^{1/4} M_{ej}^{1/4}}   t ^{5/4}
\label{R0}
\ee
In this case the break out at $R= V_{ej} t$ occurs at time and distance
\ba 
t_0 =\frac{3 }{16} \frac {c M_{ej} V_{ej}}{L_j} = 1.01 {\rm sec} M_{ej,-2} L_{51} ^{-1} \left ( \frac{\beta_{ej}} {0.3} \right)
\label{t0}
\\
R_0= V_{ej} t_0 = \frac{3 }{16} \frac {c M_{ej} V_{ej}^2}{L_j}= 9 \times 10^9 {\rm cm} M_{ej,-2} L_{51} ^{-1} \left ( \frac{\beta_{ej}} {0.3} \right)^2
\label{td}
\ea
Time $t_0$ is a typical time of the  jet-ejecta interaction.

Relations (\ref{t0}-\ref{td}) provide clear simple estimates for the break-out moment  and radius.
For example, relation (\ref{t0}) explains the result of \cite{2019ApJ...876..139G}, their Fig. 5, which shows  longer break-out times 
in the faster-expanding ejecta. 

For finite delay  times $t_d$  the edge of the ejecta is at  $R=V_{ej} (t+t_d) = \beta_0 c (t+t_d)$ (time $t$ is counted from the jet's initiation, not from the initial explosion).
Thus, the delay between the expulsion of the ejecta and turning-on of the jet  is important for $t_d >  t_0$.

At a given moment the shock velocity is 
\be
V_s = \partial_t R = 
\frac{\beta_0^{3/4} }{2\times  3^{1/4}}  \left( \frac{ 5 \sqrt{ t+t_d}- 4 \sqrt{t_d} }{ \sqrt{\sqrt{ t+t_d}- \sqrt{ t_d}} } \right) \left( \frac{\sqrt{c} L_j^{1/4} }{M_{ej}^{1/4}} \right)
\ee

At the  location  of the shock the upstream velocity is 
\be
v_1= \frac{R}{t+t_d} =\frac{2 \beta_0^{3/4} }{3^{1/4}} \sqrt{\sqrt{ t+t_d}- \sqrt{ t_d}} \frac{\sqrt{c} L_j^{1/4} }{M_{ej}^{1/4}}
\ee
So that the Mach number is
\be
M= \frac{V_s}{v_1} =  \frac{ 5 \sqrt{ t+t_d}- 4 \sqrt{t_d} } { 4  (\sqrt{ t+t_d}-  \sqrt{t_d} )  } \rightarrow \frac{5}{4}
\ee
where the last relation assumes $t_d=0$. For longer delays the Mach number is larger.  Thus, the forward shock is mild regardless of the jet power. (Stronger jets  quickly drive the shock further out, where the velocity of the ejecta is larger.)

Let's  renormalize  time by
\ba &&
t_0 =  \frac{3 }{16} \frac {c M_{ej} \beta_0  c}{L_j} 
\nn &&
\hat{t} =  \frac{t}{t_0}
\label{norm}
\ea
and the  radius of the jet-blown cavity  by the overall radius of the ejecta:
\be
 \hat{R} =\frac{R}{ \beta_0  c (t +t_d) }< 1
\ee
Value of $\hat{R}$ corresponds to the relative value of the jet-blown bubble with respect to the overall  radius of the ejecta.

\begin{figure}[h!]
\centering
\includegraphics[width=.99\textwidth]{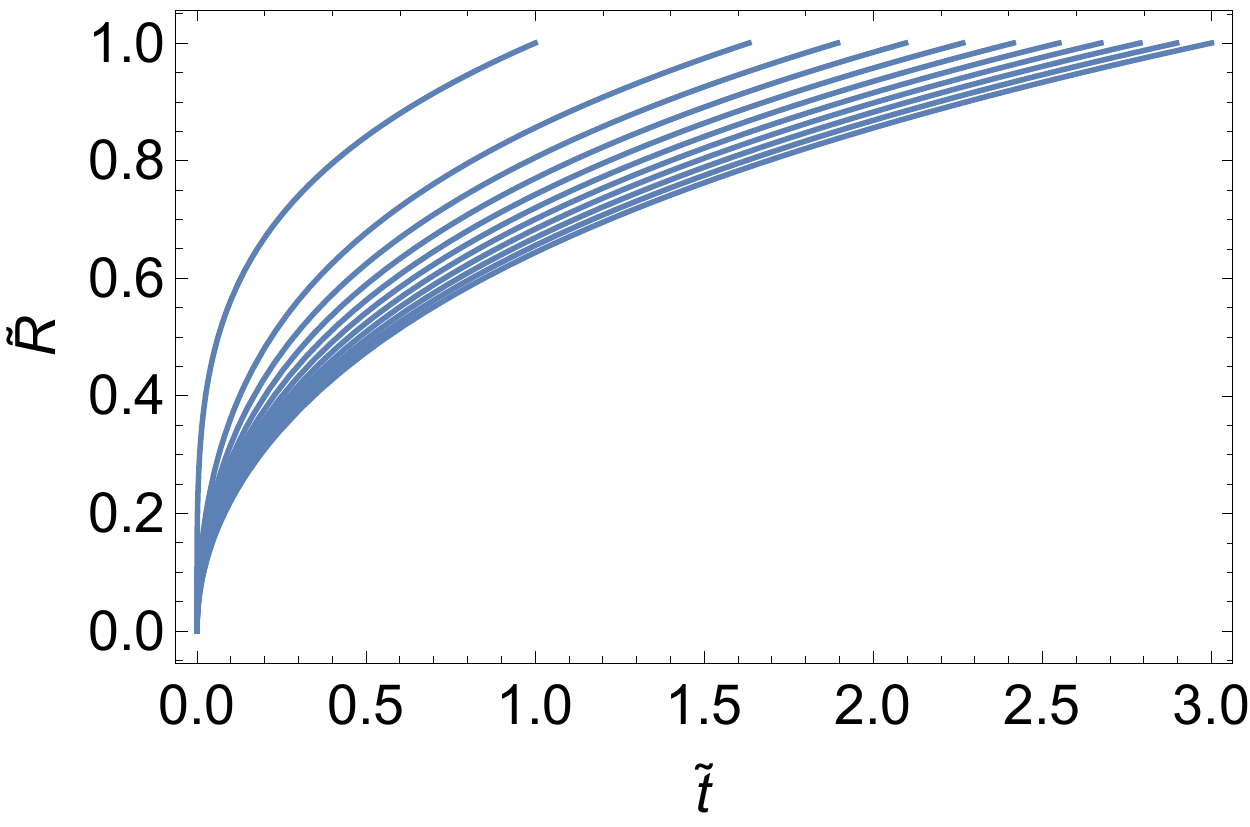}
\caption{Radius of the jet-driven bubble as a function of  time for constant jet power, normalized to the total radius of the ejecta. Different curves correspond to different delay times, $ \hat{t}_d =0, .1, ... 1$ (top to bottom).  Moments when  $\hat{R}=1$ (occurring at  $\hat{t}_{br} = 1 + 2 \sqrt{ \hat{t}_d}$) correspond to jet break-out. Corresponding physical distance is given by Eq. (\protect\ref{RRbr}). }
\label{aoft} 
\end{figure}

In dimensionless units
\be
\hat{R} =\sqrt{ \sqrt{\hat{t}+ \hat{t}_d} - \sqrt{\hat{t}_d}  } 
\approx
\left\{ 
\begin{array}{cc}
\hat{t}^{1/4}, &  \hat{t}_d \rightarrow 0
\\
\frac{1}{\sqrt{2} } \frac{\hat{t}^{1/2}}{\hat{t}_d^{1/4}}&  \hat{t}_d \rightarrow \infty
\end{array}
\right.
,
\label{Rdim}
\ee
see Fig. \ref{aoft}.

The  break out is at  $\hat{R} =1$; it occurs at time $\hat{t}_{br}$ and  physical distance $R_{br}$
\ba &&
\hat{t}_{br} = 1 + 2 \sqrt{ \hat{t}_d}
\nn && 
\frac{R_{br}}{R_0} =  1 + 2 \sqrt{ \frac{{t}_d}{t_0}}
\label{RRbr}
\ea

At the break out the energy deposited by the jet satisfies
\be
E_j=  L_j t_{br}=   \frac{5 (1+  2 \sqrt{{t}_d/t_0 })}{8 \beta_0}  E_{ej}
\label{direct}
\ee
(since we are in a regime $\beta_0\ll 1$ we neglect the difference between the emitted and absorbed power).  Thus, in order to break out   during  jet activity even with small delay $\hat{t}_d\ll 1$, the required total jet energy is fairly large, $E_j\approx  E_{ej}/\beta_0$.

\section{Late  break-outs}
\label{snowplow}

Suppose next that   the central engine stops producing a jet before the head of the jet breaks out from the ejecta.  At this moment the jet has swept some mass and moment from the ejecta, as well as deposited momentum in the shocked ejecta shell.  After the switching-off of the engine the shocked ejecta shell starts to relax as the reverse shock in the jet propagates back to the origin. This takes relatively long time, so for the times scales of interest, few seconds, the system never reaches fully relaxed Sedov stage. 

Previously \cite{2011MNRAS.411.2054L}   discussed evolution of a non-spherical shock in a steep density gradient  of expanding envelope of the exploding star (in application to long GRBs),   taking into account the sideways expansion of the jet-driven bubble.  The  approach of 
 \cite{2011MNRAS.411.2054L} follows the original  Kompaneets pressure-equilibrating  prescription \citep{1960SPhD....5...46K}. 
\cite{2019MNRAS.489.2844I} followed this procedure in details. 

It is not clear if   pressure-equilibrating assumption is applicable to the jet propagating in short GRBs. Typically, pressure equilibration takes few dynamical times.  In case of long GRBs, the velocity of the heads is expected to be highly sub-relativistic, allowing for pressure equilibration.
In the present case both the expansion velocity and the sound speed are mildly relativistic, so  pressure equilibrium within the cocoon is likely not reached fast enough. 

A related approximation that can be used is the  snowplow: we assume that  the shell propagates in momentum-conserving (snowplow) state. After the engine stops, the 
 momentum of the swept up shell  with mass $M_s$ changes due to the    swept-up momentum $P_s$:
\ba && 
\partial_t (M_s \partial _t R) = \partial_t P_s
\nn &&
M_s = 4\pi \int _0^R  \rho_{ej}  r^2 dr= M_{ej} \frac{R^3}{( c \beta_0 (t+t_d))^3}
\nn &&
P_s= 4 \pi \int_0^R \rho_{ej} \frac{r}{t+t_d} r^2 dr= \frac{3}{4} \frac{M_{ej} R^4}{ (t+t_d)^4 ( c \beta_0)^3}
\nn &&
\rho_{ej} (t)= \frac{3}{4\pi} \frac{M_{ej}}{( c \beta_0 (t+t_d))^3}
\label{33}
\ea

Eq. (\ref{33}) becomes
\be
R \left(\left(t_d+t\right){}^2 \partial _t ^2{R}+3 R\right)+3 \left(t_d+t\right){}^2
   \left(\partial _t {R}\right)^2-6 R \left(t_d+t\right)\partial _t {R}= 0
   \ee
which 
has a solution
\be
R_{after} = (t+t_d)^{3/4} \left( C_1 t   +C_2 \right)^{1/4} 
\ee
where  coefficient $C_1$ and $C_2$ are integration constants.
Converting to dimensionless  units
\be
\hat{R}_{after} =\frac{(\hat{C}_1\hat{t}   +\hat{C}_2)^{1/4}}{ (\hat{t}+\hat{t}_d)^{1/4}}
\label{R1}
\ee
Coefficients  $\hat{C}_1$ and $\hat{C}_2$  in Eq. (\ref{R1}) 
  can be derived from the condition that the radii immediately before the switch-off, Eq. (\ref{Rdim}) and after, Eq (\ref{R1}), match,
\ba &&
\hat{C}_1= 3 \hat{t}_d + 2 \hat{t}_{off} - \sqrt{ \hat{t}_d (\hat{t}_d+\hat{t}_{off})}
\nn &&
\hat{C}_2= 2 \hat{t}_d ^2 - \hat{t}_{off}^2+ (\hat{t}_{off}- 2 \hat{t}_d )  \sqrt{ \hat{t}_d (\hat{t}_d+\hat{t}_{off})}
\label{C1C2}
\ea

 \begin{figure}[h!]
\centering
\includegraphics[width=.99\textwidth]{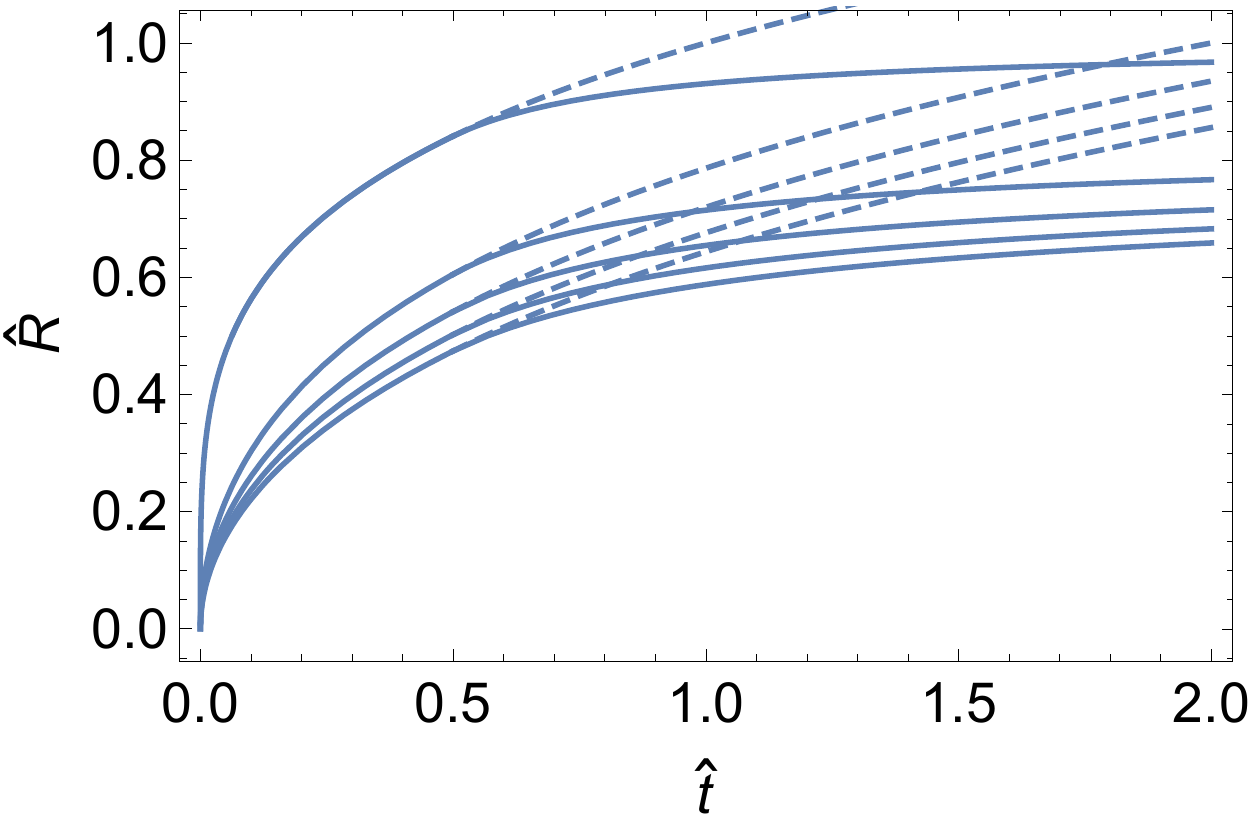}
\caption{Relative radius of the jet-driven bubble as a function of  time. At $ \hat{t}_{off}=1/2$ the jet is turned off. Different curves correspond to  $\hat{t}_d =0, 0.25,0.5,0.75, 1$ (top to bottom). After the jet is turned off  the shock quickly reaches the coasting phase. Dashed lines for $\hat{t}\geq \hat{t}_{off} =1/2$ correspond to the case without jet termination}
\label{aoft1} 
\end{figure}

Asymptotically, $\hat{t}\rightarrow \infty$, the shell is just advected with the flow,
\be
\hat{R}_{after} \approx f_1(\hat{t}_{d} ,\hat{t}_{off})   + f_2(\hat{t}_{d} ,\hat{t}_{off}) \frac{1}{\hat{t}}
\ee
where $f_1  $  and $f_2  $ are some function of $\hat{t}_{d} $ and $\hat{t}_{off}$. 
Thus, if the jet did not break out during the active phase,  after the termination of the injection the head of the jet  advances (with respect to the ejecta's flow)  very slowly $\propto  {\hat{t}}^{-1}$, see Fig. \ref{aoft1}. Particular values of coefficients  (\ref{C1C2}) are not  relevant,  and highly dependent on the assumed analytical  approximations. 

Thus, we find that  the shock approaches the edge of the ejecta very slowly  as a decreasing power-law,  $\Delta \hat{R} \propto - 1/ t$. As a result, we conclude that any analytical approximation to the late break out is likely to depend on the  subtle assumed details. Numerical models are more reliable in this case  \citep{2005A&A...436..273A,2018ApJ...866....3D,2018MNRAS.479..588G,2019arXiv190905867H}.

\section{Conclusion}

In this paper we considered conditions for the break-out of the BH-launched jet from the envelope of the ejecta material in \NS\  mergers. We provide clear  simples estimates (\ref{RRbr}) for the jet break times and conditions. Simplicity or relations (\ref{RRbr}) is quite remarkable if compared with previous similar previous approaches.
 The direct break out condition  (\ref{direct})  requires that the  total energy in jet should exceed the energy in the ejecta by at least a factor $1/\beta_0 \geq 1$.  If there is a substantial delay between the ejecta's launch and the formation of the jet, the requirement on the jet's total energy  mildly increases. Late  break outs occur only if the head of the jet relatively close to  the edge of the ejecta at the moment when the engine shuts off.

\section*{Acknowledgments}
This work had been supported by 
NASA grant 80NSSC17K0757 and  NSF grants 10001562 and 10001521. I would like to thank  Maxim Barkov and Paul  Duffell for discussions and comments on the manuscript.  


 \bibliographystyle{apj}
\bibliography{/Users/maxim/Home/Research/BibTex}

\appendix

\section{Relativistic Kompaneets equation}
\label{relatKomp}

Consider interaction of two cold flows with rest frame enthalpies $w_{1,2}$  each moving with \Lf\ $\Gamma_{1,2}$. 
In the  Kompaneets approximation (in  a sense that only momentum conservation is used), the internal dynamics in the shock regions is neglected. Then in the rest frame of the contact discontinuity
\be
u_1^2 w_1 = u_2^2 w_2  
\ee
where $u_{1,2}$ are momenta of fluids.

Let $v_w$ be the wind velocity in the lab frame, $v_{ex}$ the upstream velocity in the lab frame, and $v_s$ is the shock velocity. Since $w_{1,2}$ are Lorentz invariant quantities, $w_1=w_w$, $w_2 = w_{ex}$.
Using relativistic transformations one finds \citep{2003MNRAS.345..575M}
\ba &&
v_s = \frac{ v_{ex} /\sqrt{\cal{L}} +v_w}{1+ 1/\sqrt{\cal{L}} }
\nn &&
{\cal{L}} = \frac{\Gamma_w ^2  w_w}{\Gamma_{ex} w_{ex}}
\ea
Using expression for the luminosity of a spherical wind
\be
L_w = 4 \pi R^2 \Gamma_w^2  w_w v_w
\ee
one finds
\be
\frac{L_w}{4\pi R^2 v_w} = w_{ex} \Gamma_{ex}^2   \frac{(v_s-v_{ex})^2}{(v_w-v_s)^2}
\label{KOmp22}
\ee
For $v_w \approx c$, $w_{ex} =\rho _{ex} c^2$  and $v_s \ll v_w$, Eq. (\ref{KOmp22}) reduces to (\ref{Komp1}) for $v_s = \partial _t R$ and $v_{ex} = R/t$.

\section{More general density profiles}
\label{profile}

For ejecta's density
\be
\rho= \frac{3-n}{4\pi} \frac{M_{ej}}{ (V_{ej} t)^3}  \left( \frac{r}{V_{ej} t} \right)^{-n}
\ee
the jet's head radius evolves according to \citep[see also][]{2011MNRAS.411.2054L}
\ba&& 
R= ( t+t_d)  \left(  (t + t_d) ^{1/2} -t_d^{1/2} \right)^{2/(4-n)} \left( \sqrt{ \frac{L_j} {c M_{ej} }}  \frac{(4-n) }{\sqrt{3-n}  } V_0 ^{(3-n)/2}\right)^{2/(4-n)} 
\nn &&
R= t^{(5-n)/(4-n)} \left( \sqrt{ \frac{L_j} {c M_{ej} }}  \frac{(4-n) }{\sqrt{3-n}  } V_0 ^{(3-n)/2}\right)^{2/(4-n)}, \, \mbox{for}\,  t_d=0
\label{RRn}
\ea
The corresponding relative distance to the break out is   $\hat{R} \propto \hat{t} ^{1/(4-n)}$.
For no delay, the breakout is at 
\be
t_0'= \frac{(3-n)}{(4-n)^2} \frac{c M_{ej} V_{ej} }{L_j}
\label{toprime}
\ee
If time is normalize to (\ref{toprime}) and distance to $ V_0 t_0'$, the break out occurs at (\ref{RRbr}) regardless of the index $n$.

\end{document}